\documentclass[useAMS,usenatbib]{mn2e}

\usepackage{graphicx}

\title[Bridging Model and Observed Stellar Spectra]{
Bridging Model and Observed Stellar Spectra
}
\author[C. Allende Prieto]{C. Allende Prieto$^{1,2}$\thanks{E-mail:
callende@iac.es}\footnotemark[1]\\
$^{1}$Instituto de Astrof\'{\i}sica de Canarias, 38205, La Laguna, Tenerife, Spain\\
$^{2}$Departamento de Astrof\'{\i}sica, Universidad de La Laguna, 38206, La Laguna, Tenerife, Spain}
\begin{document}

\date{Accepted 2010 September 14.  Received 2010 September 13; in original form 2010 August 20}

\pagerange{\pageref{firstpage}--\pageref{lastpage}} \pubyear{2002}

\maketitle

\label{firstpage}

\begin{abstract}

Accurate model stellar fluxes are key for the analysis of observations of 
individual stars or stellar populations.  Model spectra differ from real
stellar spectra due to limitations of the input physical data and 
 adopted simplifications, but can be empirically 
calibrated to maximise their resemblance to actual stellar spectra.
I describe a least-squares procedure of general use 
and test it on the MILES library.

\end{abstract}


\section{Introduction}
\label{intro}

Population synthesis models build on both the theory of stellar structure
and evolution, and stellar atmospheres. The spectrum of a given stellar
population, characterised by a single age and particular chemical composition, 
is the composite light from stars with a wide 
range of mass in proportions consistent with an adopted initial mass 
function. In the calculations, the spectra of individual stars are 
sampled from a library, which is either obtained from radiative transfer
calculations in model atmospheres (see, e.g.,
Schiavon, Barbuy \& Bruzual 2000; Gonz\'alez-Delgado
 et al. 2005; Coelho et al. 2007) 
or  from observations (see, e.g.,
Bruzual \& Charlot 2003; 
Le Borgne et al. 2004; Vazdekis et al. 2010). 
Such libraries are also used to classify or parameterise 
individual stellar spectra.

The use of libraries of observed spectra has the clear advantage that
the fundamental building blocks for the models correspond to real
stars, and bypass all  approximations involved in
computing model atmospheres and synthetic spectra, including the
need for accurate atomic and molecular data.
On the other hand, using real stars has the limitation that they
are sampled from the solar neighbourhood, and therefore are limited
to what nature provides in this small region of the universe, and
their distribution in the parameter space is far from 
uniform, sampling some regions of interest quite sparsely.

An interesting option is to combine the best of both worlds,
using a complete and fairly well-sampled grid of model spectra,
and tying it to a library of observed spectra.
Such a calibration can be formulated as a
least-squares problem, where we allow smooth corrections 
to the model grid in order to match as close as possible the
available observations. In this paper we develop a procedure to 
this end, and apply it to the MILES spectral library
described by S\'anchez-Bl\'azquez et al (2006) and
Cenarro et al. (2007).

\section{Procedure}
\label{procedure}

We consider modelling the differences between synthetic spectra of
stars and the available observations with a smooth function. 
In particular, we propose to evaluate at each frequency 
the ratio between the observed spectra ($O^{\lambda}$) 
and the models ($M^{\lambda}$), and 
approximate its dependence on the stellar parameters by
polynomials. We consider spectral energy distributions, i.e.
spectra that preserve the shape of the continuum, although
the problem could be formulated as well with continuum-normalised
spectra.

We  assume that there are only three atmospheric
parameters characterising the model spectra: the surface temperature
($T_{\rm eff}$) and gravity ($\log g$), and the overall metallicity
([Fe/H]\footnote{[Fe/H] = $\log \left( N_{Fe}/N_H \right) - \log \left( N_{Fe}/N_H \right)_{\odot}$}). 
These parameters are transformed into normalised
quantities for convenience. For $T_{\rm eff}$ we define
\begin{equation}
X \equiv  (T_{\rm eff}-min(T_{\rm eff}))/(max(T_{\rm eff})-min(T_{\rm eff})),
\end{equation}
\noindent and similar transformations are applied to $\log g$ and [Fe/H] to 
define the variables $Y$ and $Z$, respectively.

Then our problem is to minimise, for each frequency\footnote{In the following we will
drop the superindex $\lambda$ for simplicity}, 
the merit function
\begin{equation}
\sum_{s} \left ( O_s/M_s - 
\sum_i \sum_j \sum_k a_{ijk} X_s^i Y_s^j Z_s^k \right)^2,
\label{chi}
\end{equation}
\noindent where the index $s$ runs through all the observed stars, and
the indices $i$, $j$, and $k$,  
run from 0 to $n$, 0 to $m$, and 0 to $h$, respectively. Thus, 
$n$, $m$, and $h$ define the orders of the polynomial in $X$, $Y$ and $Z$,
respectively. The model fluxes $M_s$ may be calculated directly for each
star, but in our case they will be obtained by interpolation in a library
(see \S \ref{synthesis}).

To find the minimum, we calculate the derivatives relative to each of
the parameters of the polynomial ($a_{\alpha\beta\gamma}$) and 
equate them to zero, arriving at 
\begin{equation}
\sum_i \sum_j \sum_k a_{ijk} \sum_s X_s^{i+\alpha} Y_s^{j+\beta} Z_s^{k+\gamma} 
= \sum_s \frac{O_s}{M_s} X_s^{\alpha} Y_s^{\beta} Z_s^{\gamma}.
\label{linear}
\end{equation}
The indices $\alpha$, $\beta$, and $\gamma$, also 
run from 0 to $n$, 0 to $m$, and 0 to $h$, respectively, 
leading to a system of $(n+1)\times(m+1)\times(h+1)$ equations.

Eq. \ref{linear} can be written in a more compact way if we map 
the sets of indices $ijk$ and $\alpha\beta\gamma$ into a 
pair of indices,
becoming 
\begin{eqnarray}
\displaystyle
\sum_{j'=0}^{(n+1)(m+1)(h+1)-1} Y_{i'j'} C_{j'} = & b_{i'}, \\ \nonumber
 i'=0,1,...,(n+1)(m+1)(h+1)-1, & \\ \nonumber
\label{linear2}
\end{eqnarray}
where 
\begin{equation}
Y_{i'j'} = \sum_s X_s^{i+\alpha} Y_s^{j+\beta} Z_s^{k+\gamma},
\end{equation}
\begin{equation}
C_{j'} = a_{ijk},
\end{equation}
\begin{equation}
b_{i'} = \sum_s \frac{O_s}{M_s} X_s^{\alpha} Y_s^{\beta} Z_s^{\gamma},
\end{equation}
with
\begin{eqnarray}
\displaystyle
i= & \lfloor j'/((m+1)(h+1)) \rfloor \\ \nonumber
j= &  j' - \lfloor i(m+1)(h+1))/(h+1) \rfloor \\ \nonumber
k = & j'-i(m+1)(h+1)-j(h+1), 
\end{eqnarray}
and similarly 
\begin{eqnarray}
\displaystyle
\alpha= & \lfloor i'/((m+1)(h+1)) \rfloor \\ \nonumber
\beta= & i' - \lfloor \alpha(m+1)(h+1))/(h+1)\rfloor \\ \nonumber
\gamma= & i'-\alpha(m+1)(h+1)-\beta(h+1),
\end{eqnarray}

\noindent where $\lfloor w \rfloor$ indicates the integer part of $w$.

Eq. 4 can be solved with any of the standard
numerical methods for solving linear system. We will use singular
value decomposition (SVD) in this paper.

\begin{figure}
\label{f1}
\centering
\includegraphics[height=60mm]{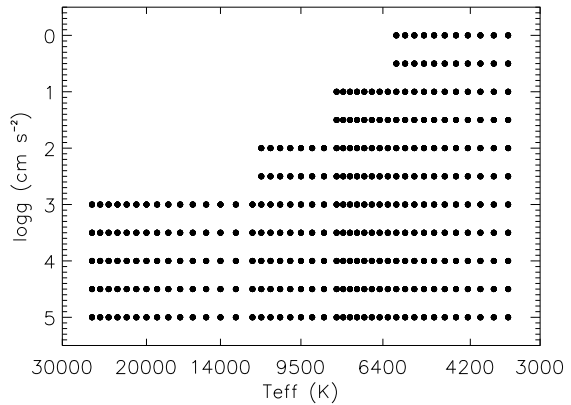}
 \caption{Coverage in surface temperature and gravity of the model spectral grids
 presented in this work. For each combination of these two parameters there 
 are 7 models covering the metallicity range between $-2.5$ and $+0.5$ dex with
 steps of 0.5 dex.}
\end{figure}

\section{Synthetic library}
\label{synthesis}

We have computed a grid of synthetic spectra covering the
wavelength range 354--716 nm with wavelength steps 
equivalent to 0.6 km s$^{-1}$. The calculations are based on
Kurucz ODFNEW model atmospheres (Castelli \& Kurucz 2004) 
and the synthesis code ASS$\epsilon$T 
(Koesterke 2009; 
Koesterke, Allende Prieto \& Lambert 2009), operated in 1D mode. 
The reference solar abundances for the synthesis are from 
Asplund, Grevesse \& Sauval (2005),
and while temperature and density are taken from the model
atmospheres, the electron density is recalculated for consistency
with the equation of state used, which includes the first 92 elements
in the period table and 338 molecules (Tsuji 1964, 1973, with some
updates). Partition functions are adopted from  Irwin (1981).

Bound-free
absorption from H, H$^-$, HeI, HeII,  and 
the first two ionisation stages of C, N, O, Na, Mg,  Al, Si, 
Ca (from the Opacity Project; see, e.g., Cunto et al. 1993) 
and Fe (from the Iron Project; Bautista 1997; Nahar 1995) is included.
Line absorption is included in detail from the atomic and molecular
(H$_2$, CH, C$_2$, CN, CO, NH, OH, MgH, SiH, and SiO) files 
compiled by Kurucz\footnote{kurucz.harvard.edu}. 
The radiative transfer calculations account 
for Rayleigh (H) and electron scattering. 
We considered stars in the ranges $3500 \leq T_{\rm eff} \leq 26,000$ K,
$0 \leq \log g \leq 5$, and $-2.5 \leq$[Fe/H]$\leq +0.5$, divided
into 5 subgrids, as shown in Table 1 and Figure 1.

\begin{table}
\label{t1}
\centering
\begin{minipage}{140mm}
\caption{Range of parameters for each subgrid.}
\begin{tabular}{@{}lrrrr@{}}
\hline
Grid & $T_{\rm eff}$ & $\log g$ & [Fe/H] & MILES stars  \\
  \#      &     (K)       &  (cm s$^{-2}$) &  (dex) &  included   \\
\hline
1       & 3500:4500    &   0:5   & $-2.5$:$+0.5$ &   190  \\
2       & 4750:6000    &   0:5   & $-2.5$:$+0.5$ &   350  \\
3       & 6250:8000    &   1:5   & $-2.5$:$+0.5$ &   124  \\
4       & 8500:11500   &   2:5   & $-2.5$:$+0.5$ &   48   \\
5       &12000:26000   &   3:5   & $-2.5$:$+0.5$ &   22   \\
\hline
\end{tabular}
\end{minipage}
\end{table}

\begin{figure*}
\centering
\includegraphics[height=140mm,angle=90]{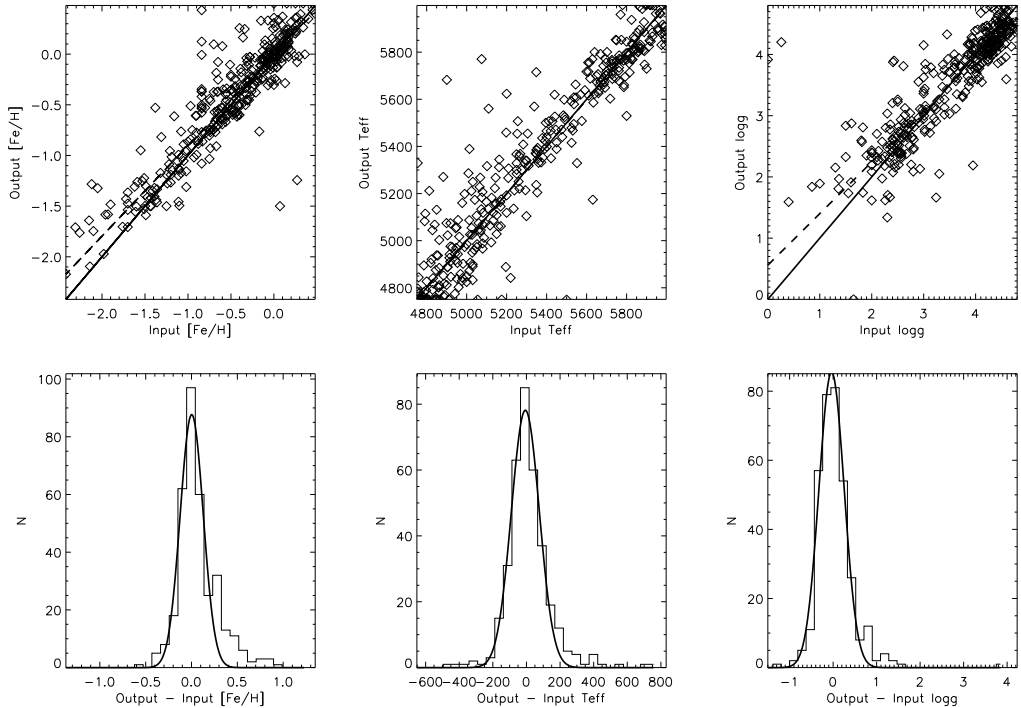}
 \caption{Comparison between the atmospheric parameters for the MILES
 stars in the range of grid \# 2 (see Table \ref{t1}). The input parameters
 are those provided with the library (Cenarro et al. 2007), and the output
 parameters those rederived in this work. In the upper panels, the solid  and
 dashed  lines are straight lines with slopes unity and determined from least-squares,
 respectively. The bottom panels show histograms of the residuals, and  
 Gaussians with parameters derived again from least-squares fitting.}
\label{f2}
\end{figure*}

The spectra were smoothed to a resolution of 0.23 nm, and resampled
in steps  of 0.09 nm, matching the resolution and wavelength scale
of the MILES library.

\section{Calibration results}
\label{cali}

We apply the fitting procedure described in \S \ref{procedure} 
to model the ratio of the observed to the synthetic spectra with
low order polynomials. The procedure is independently applied
to each of the subgrids in Table 1. We tested with
various combinations of first, second, and third order polynomials
for each parameter, finding that second-order or higher dependencies
for some parameters, even if improved agreement with the observations
for the stars under consideration, led to unphysical shapes in 
poorly constrained regions of the parameter space.

After some experimentation, the adopted polynomials were
second order in temperature ($n=2$), zero-th order in surface 
gravity ($m=0$),
and up to first order in metallicity ($h=1$ for grids \#1, 2 and 3, 
but  $h=0$ for grids \#4 and 5). The polynomial model 
for the ratio of  observed and synthetic spectra was then 
used to correct the synthetic grid, and new synthetic spectra
for the parameters of the MILES stars were derived by quadratic
Bezier interpolation. As expected the corrections   
tightened, made more symmetric, and centred closer to zero 
the distributions of residuals. 


Particularly large residuals remain after the correction for some stars
with cool temperatures and low gravities, suggesting that these stars
are somewhat singular, or that their assigned parameters are wrong.
To improve consistency, we redetermined the atmospheric parameters
for all stars by using  the most recent version of the 
optimisation code discussed by  
Allende Prieto et al. (2004, 2006, 2008, 2009),
and exclude a few stars which could not be fit reasonably well.
Fig. 2 illustrates the comparison between the old and
the rederived parameters for grid \# 2.  Robust estimates 
of the mean and the standard deviation (half of
the width of the distribution after discarding 15.85 \% of the sample
on each end) between the MILES parameters and the redeterminations
are given in Table 2, which also includes the numbers of surviving 
library stars within each of the subgrids.

\begin{figure}
\centering
\includegraphics[height=120mm]{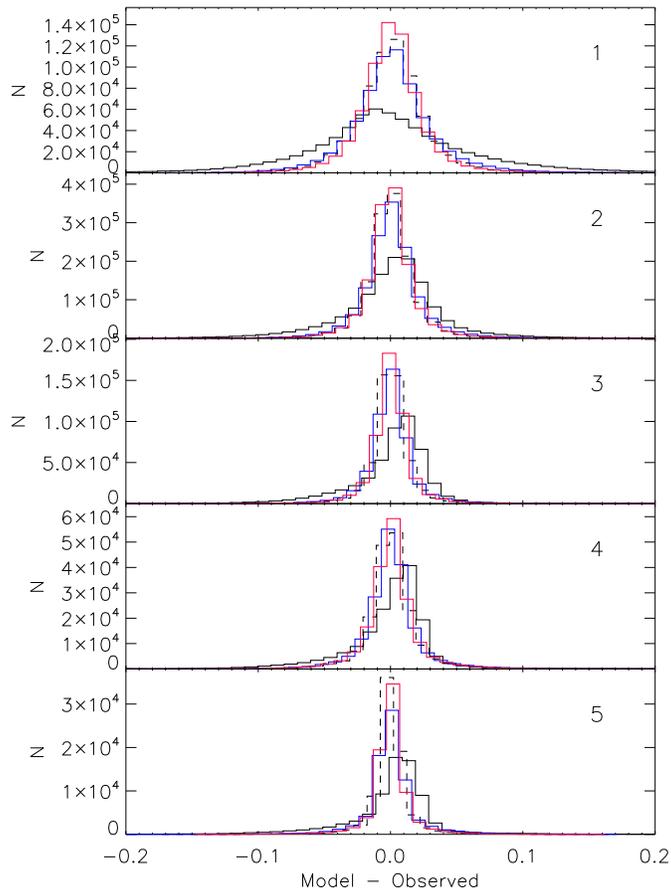}
 \caption{Residuals for all the stars in each subgrid, and all wavelengths, 
 using the original synthetic libraries (solid black line),  the
 corrections using  $n=m=h=1$ (solid dashed line), those for $n=2$, $m=0$, $h=1$ (red),
 and those for $n=2$, $m=0$, $h=0$ (blue).}
\label{f3}
\end{figure}

\begin{table}
\label{t2}
\centering
\begin{minipage}{140mm}
\caption{Standard deviation between the original and updated par.}
\begin{tabular}{@{}lrrrr@{}}
\hline
Grid & $T_{\rm eff}$ & $\log g$ & [Fe/H] & MILES stars  \\
        &     (K)       &  (cm s$^2$) &  (dex) &  surviving   \\
\hline
1       & 94    &   0.48   & 0.26 &   166  \\
2       & 98    &   0.31   & 0.19 &   337  \\
3       & 128    &  0.19   & 0.17 &   118  \\
4       & 616   &   0.30   & 0.51 &   45   \\
5       & 783   &   0.19   & 0.39 &   19   \\
\hline
\end{tabular}
\end{minipage}
\end{table}

The parameters provided with MILES (Cenarro et al. 2007) have been compiled from 
the literature, and subsequently homogenised by identifying and removing 
systematics across data sets (Cenarro et al. 2001).
The metallicities in this compilation are mainly from high-resolution
studies, and therefore reflect measurements from iron lines. 
Our model spectra have simply solar scaled metal abundances and,
interestingly, if the corrections introduced are independent of
[Fe/H], i.e. $h=0$, our rederived metallicities are systematically
higher than those in the MILES catalogue for metal-poor stars. 
With $h>0$, such a trend disappears, as the corrections partially 
account for the relative strengthening of the features produced 
by $\alpha-$elements relative to iron.

The polynomial fitting using the algorithm described in Section \ref{procedure}
was then repeated using the original grid of synthetic spectra and
the rederived atmospheric parameters. New model fluxes for the MILES stars
were recalculated by quadratic Bezier interpolation, and the residuals between
these and the observed fluxes are displayed in Fig. 3.

\begin{figure}
\includegraphics[height=120mm]{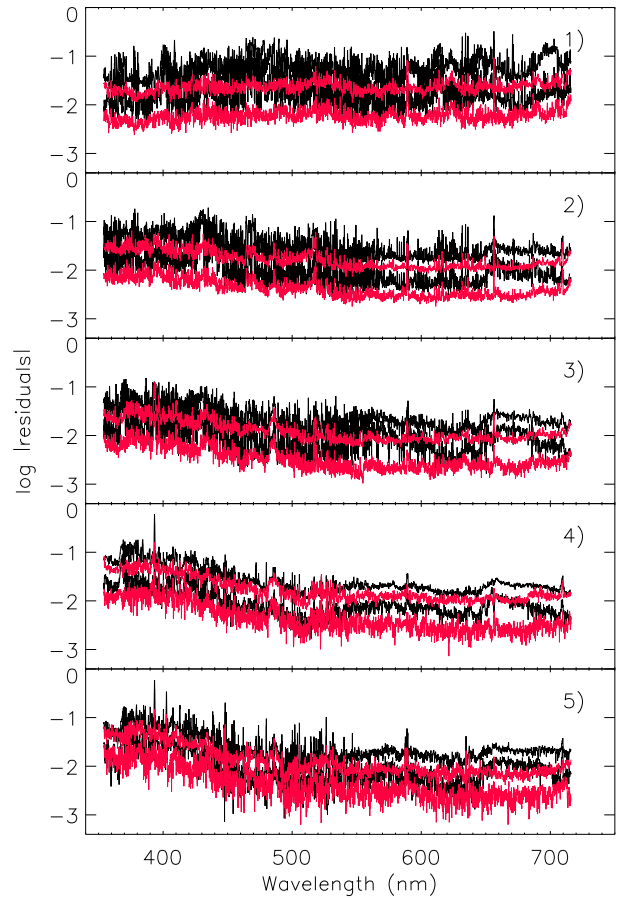}
 \caption{The lines show the 25\% and 75 \% percentiles of the
 decimal logarithm of the absolute value of the 
 residuals between the model and the observed  spectra for the
 original theoretical library (black) and the 2nd-order corrected
 (1st-order for the warmest subgrid, shown in the bottom panel) 
 library (red).}
\label{f4}
\end{figure}

Fig. 4 shows the 25\% and the 75\% percentiles of the absolute
value of the residuals 
for each of the subgrids as a function of wavelength. The black
lines are the original results for the theoretical library, and
the red lines correspond to the same percentiles for the corrected library.
Not only the residuals are decreased for essentially all the
parameter space, but they are also smoother with wavelength
for cool spectral types.

\begin{figure*}
\centering
\includegraphics[height=180mm,angle=90]{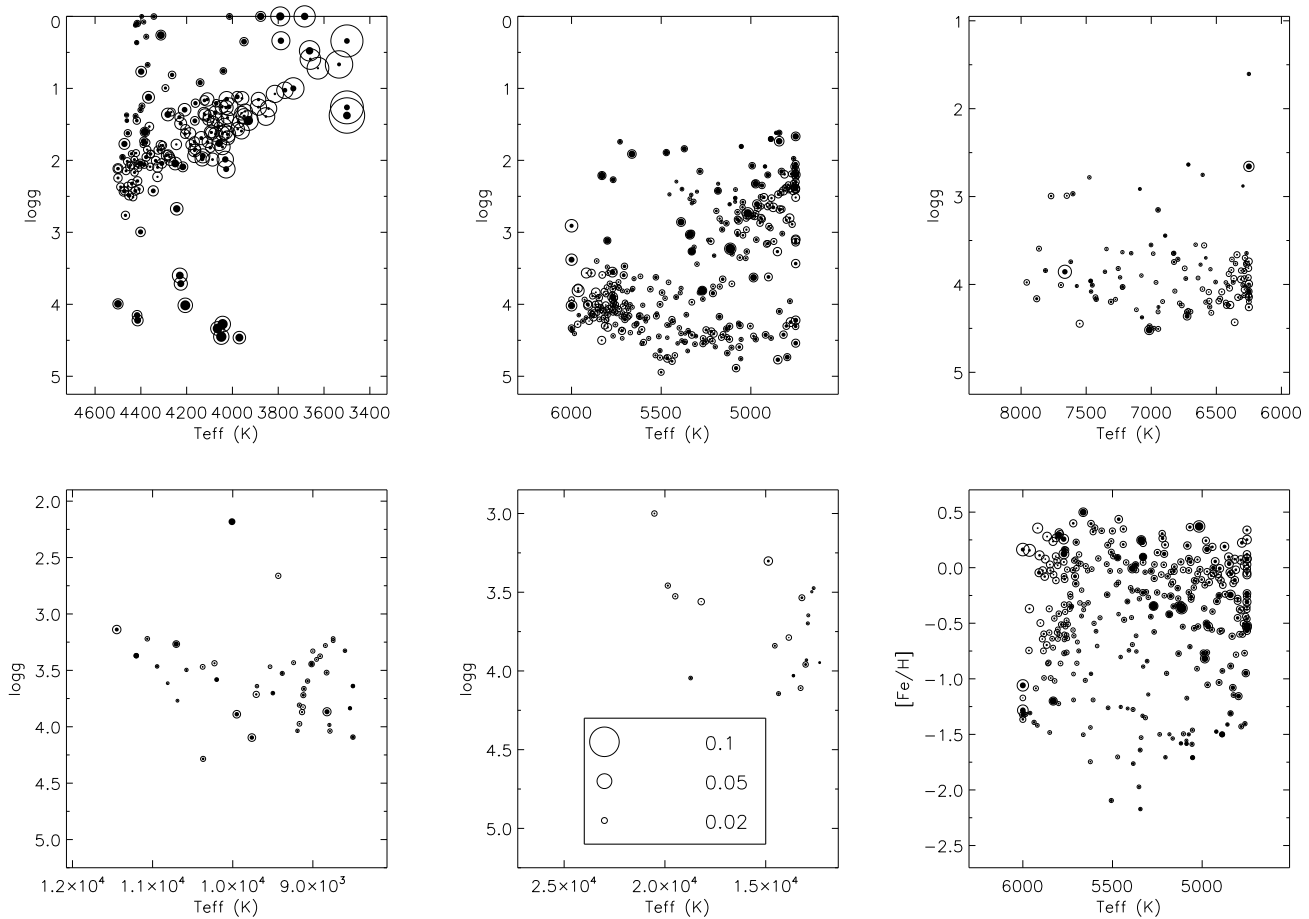}
 \caption{Absolute value of the residuals for individual stars 
 in each of the grids. The open circles correspond to the original
 grid of synthetic spectra, and the filled circles to the grid 
 corrected with first-order polynomials. The residuals are projected
 onto the $T_{\rm eff}$--$\log g$ plane for all five grids, and
 also for the $T_{\rm eff}$--[Fe/H] plane for grid \#2 in the bottom-left
 panel.}
\label{f5}
\end{figure*}

Fig. \ref{f5} illustrates the same results from  a 
different perspective.
The absolute value of the residuals are now shown for individual
stars for each subgrid. Each star has two data points in the
plot, at the same location, differing only on the size and type 
of symbol: open circles correspond to the original synthetic grid,
and filled circles to the second-order corrected grid. The 
corrected library is in generally closer to the observations, 
but not in all cases, as permitted by the least-squares approach.

To ensure that we are not overfitting the data, 
the polynomial modelling was repeated using only 2/3 of the stars, 
examining the changes in the residuals between the 
corrected grid and the observations for the remainder of the
stars. This test showed also a significant reduction in the
residuals, in line with the results obtained when fitting the
complete sample.

\section{External check}

A sanity check of the empirical spectral energy distributions
can be done by using stars with reliable spectrophotometric 
calibrations. We have chosen three key stars: the Sun,
BD $+17$ 4718,  and Vega. Our choice of
atmospheric parameters for these stars, compiled from the literature, 
are shown in Table 3. The data are from the calibration archive 
for HST, calspec (Bohlin 2008, 2010, and references therein); 
they come from HST observations for all
the stars but the Sun, which are from the compilation by Colina, Bohlin, 
\& Castelli (1996). We have dereddened the spectrum of BD $+17$ 4718
according to the prescription by Fitzpatrick \& Masa (Fitzpatrick 1999), 
using $A_V/E(B-V) = 3.1$, consistent with the corrections applied to MILES.

\begin{table*}
\label{t3}
\centering
\begin{minipage}{140mm}
\caption{Testing stars.}
\begin{tabular}{@{}lrrrrr@{}}
\hline
Star & $T_{\rm eff}$ & $\log g$ & [Fe/H] & E(B-V) & References  \\
        &     (K)       &  (cm s$^2$) &  (dex) &  (mag) &    \\
\hline
Sun            & 5777    &   4.44   & $ 0.0$  &   0.00  & Stix  (2004) \\
BD $+17 4708$  & 6141    &   3.87   & $-1.7$  &   0.01  & Ram\'{\i}rez et al. (2006) \\
Vega           & 9620    &   3.98   & $-0.7$  &   0.00  & Garc\'{\i}a-Gil et al. (2005)   \\
\hline
\end{tabular}
\end{minipage}
\end{table*}

We warn the readers that parameters adopted for BD $+17$ 4708 and 
Vega rely on the HST spectrophotometry for these stars
and model fluxes based on Kurucz atmospheres, and therefore
the excellent agreement found for our purely theoretical fluxes
is somewhat facilitated. The test is 
only completely fair for the Sun, as this is
the only star in this group with parameters that are truly independent
from model atmospheres (and have negligible errors), 
but the others are included as they may 
help to confirm the conclusions found from the examination of the
solar case.

Fig. 6Ê compares the predicted spectral energy distributions,
obtained by linear interpolation in a) our purely theoretical
grid of model spectra (red line in left-hand panels), and b) 
our empirically calibrated grid (red line in right-hand panels).
The  interpolated spectra have also been smoothed to approximate
the lower resolution of the HST spectrophotometry, with a FWHM resolving
power $\lambda/\delta\lambda \sim 1000$. Overall, the calibrated 
grid performs slightly better than
the purely theoretical grid in some regions, but the opposite 
is true in others. 

This result suggests that there may be
some systematic differences between the flux scale adopted for 
HST calibration and that of the MILES library, and the 
corrections derived from MILES may not be widely applicable.
To keep a perspective, we note that systematic and random errors 
in the MILES fluxes were estimated to be about 2 and 3 \%, respectively,
from the comparison with photometry ($B-V$) from the Lausanne data base
(Mermilliod et al. 1997). A reduction in effective temperature for 
the solar model by 150 K -- which is a typical error in this quantity
to be expected for most stars (not the Sun)-- would be enough to 
bridge the 5\% gap in the red between the solar fluxes and those
from the empirically corrected library.

\begin{figure}
\centering
\includegraphics[height=120mm,angle=0]{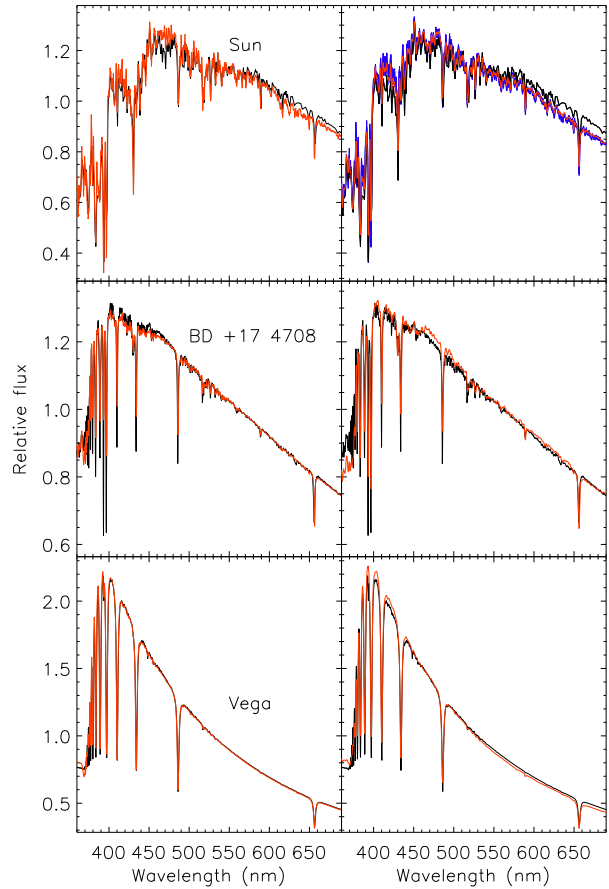}
 \caption{Observed (CALSPEC) spectrophotometry for three flux {\it standards}
 (Sun, BD $+17 4707$, and Vega; in black) compared with spectra interpolated from our grids
 (red lines):
 left-hand side graphs for the purely theoretical grid, and right-hand side for 
 the MILES-calibrated grid. In the top right-hand panel, several spectra of solar-like
 stars from the MILES library are shown in blue.}
\label{f6}
\end{figure}

It should be emphasised that an empirical calibration, such as  
the procedure described in this paper, will equally erase 
discrepancies between models and observations due to model
deficiencies and due to systematic errors in the observations,
and the latter are highly dependent on the source of the data.
In the upper-right panel of Fig. \ref{f6}, we also show in blue
the spectra of three G2V stars included in the MILES library:
HD 10307, HD 84737, and HD 13043. As expected, their fluxes are consistent
with the spectrum interpolated in the corrected 
library\footnote{A fourth G2V star HD 76151 in the library, however, shows 
fluxes that more closer to the reference solar observations.}. 
An  interpolated spectrum  for the solar parameters
from neighbouring stars using the tool available at the 
MILES web 
site\footnote{http://www.iac.es/proyecto/miles/pages/webtools/star-by-parameters.php}
leads to the same conclusion. 

It is interesting to highlight the poor matching between the
original library and the observed strength of the Ca II H and K
lines in BD $+17$ 4708. This mismatch, which can be directly
ascribed to the use of solar abundance ratios in the models,
and hence the neglect of the enhancement in $\alpha$ elements 
typically observed in metal-poor halo stars, is reduced
somewhat after the polynomial corrections have been introduced.




\section{Summary and conclusions}

We present a general strategy to calibrate a set of model stellar fluxes 
using observations for a network a flux standards. 
At each frequency, the ratio of model and observed fluxes is fit
by least-squares to a polynomial that depends on the atmospheric 
parameters. We test this scheme on the MILES stellar library, 
with satisfactory results: second-order corrections on effective
temperature, zero-th order corrections on surface gravity, and
zero-th or first-order corrections on metallicity perform well
and improve significantly the agreement between the model grid
and the observations, resulting in  tighter and more symmetric
distributions of residuals.

Seeking an external assessment of the performance of the
corrections, we take a close look at the spectra of 
three well-known flux standards  (see Bohlin 2007 and
references therein).
The result of this test is a much modest than expected improvement,
showing that the purely theoretical fluxes perform 
better than those corrected for some wavelengths and stars.
This suggests that the HST fluxes and those in the
MILES library are not fully compatible, but further 
investigation is warranted.

The HST/STIS Next Generation Spectral Library (Gregg et al. 2005) 
has been recently released
as part of the high-level products in the Multi-mission
Archive at the Space Telescope (MAST). This library 
is based on observations made in cycles 10 through 13
and includes nearly 400 stars. This number
is smaller than those in the MILES library and the resolution
of the spectra is also lower, but the 
spectral coverage is wider. An application of the method
presented here to this library, or if compatible
with MILES, to the combination of the two libraries would
be useful and is planned for the future. 

\section*{Acknowledgements}

I thank Alexandre Vazdekis and Jes\'us Falc\'on-Barroso for inspiration
for this work and stimulating discussions.  
Ignacio Ferreras contributed a number of suggestions that improved
the paper. Critical contributions to the spectral synthesis calculations by
Paul Barklem, Manuel Bautista, Ivan Hubeny, Lars Koesterke and Sultana Nahar are 
gratefully acknowledged.


\begin{thebibliography}{99}

\bibitem[Allende Prieto et al.(2009)]{2009MNRAS.396..759A} Allende Prieto, 
C., Hubeny, I., \& Smith, J.~A.\ 2009, MNRAS, 396, 759 


\bibitem[Allende Prieto et al.(2008)]{2008AJ....136.2070A} Allende Prieto, 
C., et al.\ 2008, AJ, 136, 2070 


\bibitem[Allende Prieto et al.(2006)]{2006ApJ...636..804A} Allende Prieto, 
C., Beers, T.~C., Wilhelm, R., Newberg, H.~J., Rockosi, C.~M., Yanny, B., 
\& Lee, Y.~S.\ 2006, ApJ, 636, 804 


\bibitem[Allende Prieto(2004)]{2004AN....325..604A} Allende Prieto, C.\ 
2004, Astronomische Nachrichten, 325, 604 

\bibitem[Asplund et al.(2005)]{2005ASPC..336...25A} Asplund, M., Grevesse, 
N., 
\& Sauval, A.~J.\ 2005, Cosmic Abundances as Records of Stellar Evolution and Nucleosynthesis, 336, 25 

\harvarditem{{Bautista}}{1997}{1997A&AS..122..167B}
{Bautista} M~A  1997, A\&AS, {\bf 122},~167--176.

\bibitem[Bohlin(2007)]{2007ASPC..364..315B} Bohlin, R.~C.\ 2007, The Future 
of Photometric, Spectrophotometric and Polarimetric Standardization, 364, 
315 

\bibitem[Bohlin(2010)]{2010AJ....139.1515B} Bohlin, R.~C.\ 2010, AJ, 139, 
1515 

\bibitem[Bohlin 
\& Cohen(2008)]{2008AJ....136.1171B} Bohlin, R.~C., \& Cohen, M.\ 2008, AJ, 136, 1171 


\bibitem[Bruzual 
\& Charlot(2003)]{2003MNRAS.344.1000B} Bruzual, G., \& Charlot, S.\ 2003, MNRAS, 344, 1000 

\bibitem[Castelli 
\& Kurucz(2004)]{2004astro.ph..5087C} Castelli, F., \& Kurucz, R.~L.\ 2004, arXiv:astro-ph/0405087 

bibitem[Cenarro et al.(2007)]{2007MNRAS.374..664C} Cenarro, A.~J., et al.\ 
2007, MNRAS, 374, 664 

\bibitem[Cenarro et al.(2001)]{2001MNRAS.326..981C} Cenarro, A.~J., Gorgas, 
J., Cardiel, N., Pedraz, S., Peletier, R.~F., 
\& Vazdekis, A.\ 2001, MNRAS, 326, 981 

bibitem[Coelho et al.(2007)]{2007MNRAS.382..498C} Coelho, P., Bruzual, G., 
Charlot, S., Weiss, A., Barbuy, B., 
\& Ferguson, J.~W.\ 2007, MNRAS, 382, 498 

\bibitem[Cunto et al.(1993)]{1993BICDS..42...39C}
{Cunto} W, {Mendoza} C, {Ochsenbein} F \harvardand\ {Zeippen} C~J  1993 {\em
  Bulletin d'Information du Centre de Donnees Stellaires}, 42,~39--+.

\bibitem[Fitzpatrick(1999)]{1999PASP..111...63F} Fitzpatrick, E.~L.\ 1999, PASP, 111, 63 

\bibitem[Garc{\'{\i}}a-Gil et al.(2005)]{2005ApJ...623..460G} 
Garc{\'{\i}}a-Gil, A., Garc{\'{\i}}a L{\'o}pez, R.~J., Allende Prieto, C., 
\& Hubeny, I.\ 2005, ApJ, 623, 460 

\bibitem[Gonz{\'a}lez Delgado et al.(2005)]{2005MNRAS.357..945G} 
Gonz{\'a}lez Delgado, R.~M., Cervi{\~n}o, M., Martins, L.~P., Leitherer, 
C., \& Hauschildt, P.~H.\ 2005, MNRAS, 357, 945 

\bibitem[Gregg et al.(2006)]{2006hstc.conf..209G} Gregg, M.~D., et al.\ 
2006, The 2005 HST Calibration Workshop: Hubble After the Transition to 
Two-Gyro Mode, 209 

\bibitem[Irwin(1981)]{1981ApJS...45..621I} Irwin, A.~W.\ 1981, ApJS, 45, 
621 

\bibitem[Koesterke(2009)]{2009AIPC.1171...73K} Koesterke, L.\ 2009, 
American Institute of Physics Conference Series, 1171, 73 

\bibitem[Koesterke et al.(2008)]{2008ApJ...680..764K} Koesterke, L., 
Allende Prieto, C., \& Lambert, D.~L.\ 2008, ApJ, 680, 764 

\bibitem[Le Borgne et 
al.(2004)]{2004A&A...425..881L} Le Borgne, D., Rocca-Volmerange, B., Prugniel, P., Lan{\c c}on, A., Fioc, M., \& Soubiran, C.\ 2004, A\&A, 425, 881 

\bibitem[Mermilliod et 
al.(1997)]{1997A&AS..124..349M} Mermilliod, J.-C., Mermilliod, M., \& Hauck, B.\ 1997, A\&AS, 124, 349 

\bibitem[Nahar(1995)]{1995A&A...293..967N}
{Nahar} S~N  1995, A\&A, 293,~967--977.

\bibitem[S{\'a}nchez-Bl{\'a}zquez et al.(2006)]{2006MNRAS.371..703S} 
S{\'a}nchez-Bl{\'a}zquez, P., et al.\ 2006, MNRAS, 371, 703 

\bibitem[Ram{\'{\i}}rez et 
al.(2006)]{2006A&A...459..613R} Ram{\'{\i}}rez, I., Allende Prieto, C., 
Redfield, S., \& Lambert, D.~L.\ 2006, A\&A, 459, 613 

\bibitem[Schiavon et al.(2000)]{2000ApJ...532..453S} Schiavon, R.~P., 
Barbuy, B., \& Bruzual A., G.\ 2000, ApJ, 532, 453 

bibitem[Stix(2004)]{2004suin.book.....S} Stix, M.\ 2004, The sun : an 
introduction, 2nd ed., by Michael Stix.~Astronomy and astrophysics library, 
Berlin: Springer, 2004.~ ISBN: 3540207414,  

\bibitem[Tsuji(1964)]{1964AnTok...9.....T} Tsuji, T.\ 1964, Annals of the 
Tokyo Astronomical Observatory, 9,  

\bibitem[Tsuji(1973)]{1973A&A....23..411T} Tsuji, T.\ 1973, A\&A, 23, 411 

\bibitem[Vazdekis et al.(2010)]{2010MNRAS.404.1639V} Vazdekis, A., 
S{\'a}nchez-Bl{\'a}zquez, P., Falc{\'o}n-Barroso, J., Cenarro, A.~J., 
Beasley, M.~A., Cardiel, N., Gorgas, J., 
\& Peletier, R.~F.\ 2010, MNRAS, 404, 1639 


\end{thebibliography}
\end{document}